\documentclass[nofootinbib,superscriptaddress,twocolumn,aps,prc,showpacs,10pt]{revtex4-1}

\usepackage{amsmath}
\usepackage{amssymb}
\usepackage{graphicx}
\usepackage{mathrsfs}
\usepackage{color}
\newcommand{\bvec}[1]{\mbox{\boldmath $#1$}}

\begin{document}
\title{Three-body model for an isoscalar spin-triplet neutron-proton pair in $^{102}{\rm Sb}$}

\author{Yusuke Tanimura} \email{tanimura@ipno.in2p3.fr}
\affiliation{Institut de Physique Nucl\'eaire, IN2P3-CNRS, Universit\'e Paris-Sud, F-91406 Orsay Cedex, France}
\author{Hiroyuki Sagawa}
\affiliation{RIKEN Nishina Center, Wako 351-0198, Japan}
\affiliation{Center for Mathematics and Physics, University of Aizu, Aizu-Wakamatsu, Fukushima 965-8560, Japan}
\date{\today}

\begin{abstract}
We discuss the effect of  neutron-proton pair in $T=0, S=1$ channel in the low-lying states of 
$^{102}{\rm Sb}={}^{100}{\rm Sn}+p+n$ nucleus in a three-body model.
To this end, we construct a  ${\rm core}+p+n$ three-body model with a model space 
 based on Skrme and relativistic mean field  calculations. 
The latter model is  found to be more realistic for the present case due to the pseudo-spin symmetry. 
It turns out that the $(L,S,T)=(0,1,0)$ coupling scheme  of valence nucleons is strongly hindered in 
$^{102}$Sb with the relativistic model because of the near degeneracy of the $g_{7/2}$ and $d_{5/2}$ 
orbitals in the valence space. 
The effect of this neutron-proton coupling is clearly seen in the charge-exchange Gamow-Teller-type transitions 
rather than in the binding energies of $T=0$ and $T=1$ states. 
\end{abstract}


\maketitle

\section{Introduction}

The pairing correlation between like nucleons, that is, the spin-singlet isovector $T=1$ pairing, 
is a crucial part of the correlations in the nuclear many-body system, which comes from 
the short-range part of nucleon-nucleon interaction. 
It is essential for understanding the various properties of open-shell nuclei, 
such as odd-even mass staggering, deformation properties, 
large energy gaps between the ground and $2^+_1$ states in the even-even systems, 
the moment of inertia of deformed nuclei. \cite{RS,Suho,FM14,Sa15}. 

In contrast, in spite of the strong attraction in $(S,T)=(1,0)$ channel which makes the deuteron bound, 
the spin-triplet  $T=0$  coupling configuration is considered to be severely hindered in 
most of the stable nuclei \cite{FM14,Sa15}. There are mainly three reasons for this. 
Firstly, the spin-orbit splitting makes it unfavorable to form a $(L,S,T)=(0,1,0)$ pair 
consisting of the coherent mixture of $(j_>)^2$ and $(j_>j_<)$ configurations, 
where $j_>=l+1/2$ and $j_<=l-1/2$ are spin-orbit partners of each other. 
Secondly, the recoupling of angular momenta from $jj-$ to
$LS-$coupling schemes  effectively quenches the pairing
interaction more in the $S = 1$ channel than in the $S = 0$
channel \cite{Sa13}.  Lastly, 
the neutron excess of the stable nuclei prevents valence protons and neutrons 
from making $(L=0,S=1)$ pairs 
because they occupy different orbitals around the Fermi energy. 

As the progress of radio-isotope beam facilities allows to access nuclei near the $N=Z$ line, 
the neutron-proton pairing has been attracting  more interest than before. 
A number of works have been devoted to this subject to investigate the 
existence of $T=0$ spin-triplet superfluidity, its competition with the $T=1$ spin-singlet one, 
and its manifestation in observables 
such as binding energies of $T=0$ and $T=1$ states, and charge-exchange or pair-transfer reactions 
(for reviews, see Refs. \cite{FM14,Sa15}). 
As a generic trend, 
a phenomenological analysis made on the systematics of the measured binding energies
in $N=Z\lesssim 30$ nuclei  shows that the $T=0$ pair condensation is not likely to exist in those nuclei \cite{Ma00}. 
On the other hand, systematic Hartree-Fock-Bogoliubov calculations performed in 
Refs. \cite{BeLu10,GBL11} demonstrated that the  isoscalar pair condensation  might  take place in 
much heavier $N\sim Z$ regions. 
However, it is still interesting and important to study some specific systems in the proximity of the $N=Z$ line, 
where the neutron-proton pair in $T=0$ channel could play a prominent role. 

In general, a condition which is expected to maximize the chance for valence nucleons to form 
$T=0$ pairs is that they lie in the orbitals on top of a $N=Z$ closed-shell core, and that both $j_>$ and 
$j_<$ orbitals with low $l$ are vacant in the valence space. 
In addition, it is favorable that the mass of nucleus is large so that the effect of 
the spin-orbit field at surface becomes relatively week \cite{BeLu10,GBL11}. 
In the present study, we focus on $^{102}{\rm Sb}={}^{100}{\rm Sn}+p+n$ system, 
whose core, $^{100}$Sn, is the largest  $N=Z$ doubly-magic nucleus located just below the proton drip line. 
In this system, the valence nucleons can occupy $d$-wave states as in $^{18}$F nucleus, where the strong 
effect of $(S,T)=(1,0)$ coupling channel is observed as the implementation of the good SU(4) symmetry \cite{Tani14}. 
Therefore, the nucleus $^{102}$Sb would be the last possible system on 
$N=Z$ line in which the $T=0$ pair coupling  between the valence nucleons may play a significant role 
although further developments of experimental facilities are needed to perform spectroscopies of 
the proton-rich nucleus.   

In order to describe the $^{102}$Sb system, we develop the ${\rm core}+p+n$ three-body model 
which is based on the self-consistent mean-field calculations. 
A similar model which combines the self-consistent mean-field and shell model technique has been employed in Ref. \cite{Ga15} 
for studying the deformation effects on the $(T=1,S=0)$ and $(T=0,S=1)$ pairs. We employ relativistic energy density functionals (EDFs)
as well as the non-relativistic ones because of  the reason to be explained in Sec. \ref{sec:edf}. 
Using this model, we can calculate the low-lying levels of the $A=102$ systems and the transition rates among them.   

The paper is organized as follows. 
In Sec. \ref{sec:model},  we describe our model.   In Sec. \ref{sec:edf},  we illustrate the characteristic features 
of the different EDFs to be used in our calculations.  
We present the results of our three-body model calculations in Sec. \ref{sec:result}. 
Section  \ref{sec:summary} is devoted for a summary of the paper.

\section{Model}\label{sec:model}

\subsection{Three-body model based on mean-field calculation}
In this work we employ a three-body model \cite{Ber91,Esb97} in which one assumes a core + two nucleons structure of the system. 
The model was originally developed for studying di-neutron pair in neutron-rich nuclei near the dripline 
\cite{Ber91,Esb97,Hagi05,Hagi14}. 
It has also been applied to proton-rich systems assuming ${\rm core} + 2p$ structure \cite{Oishi10,Oishi14}. 
In Ref. \cite{Tani14}, we have extended the model to ${\rm core} + p + n$ systems to study 
neutron-proton pair  in $T=0$ channel in $N=Z$ nuclei. 
The three-body Hamiltonian for ${\rm core} + p + n$ system is given by 
\begin{eqnarray}
H &= & \frac{\bvec p_p^2}{2m}+\frac{\bvec p_n^2}{2m}+V_{pC}+V_{nC}
\nonumber\\
&&
+V_{\rm res}(p,n)+\frac{(\bvec p_p+\bvec p_n)^2}{2A_Cm}, 
\label{eq:3bWS}
\end{eqnarray}
where the first and second terms are the kinetic energies of the valence nucleons, and the 
third and the fourth terms are core-nucleon mean-field potentials, 
the fifth term is the residual interaction between the valence nucleons, and the last term is the recoil kinetic energy 
of the core. 
The core-nucleon potentials $V_{nC}$ (and $V_{pC}$) were given in Ref. \cite{Tani14} as a 
spherical Woods-Saxon potential with central and spin-orbit parts (plus the Coulomb potential). 

In the present case where we have $^{100}$Sn as the core, it is difficult to determine the parameters of the 
Woods-Saxon potential since the data are quite poor for this nucleus. 
Thus we take instead a self-consistent mean-field Hamiltonian for the single-particle part, 
\begin{eqnarray}
H = h_{\rm HF}(p)+h_{\rm HF}(n)+V_{\rm res}(p,n), 
\end{eqnarray}
where $h_{\rm HF}$ is the single-particle Hamiltonian obtained with a Hartree-Fock calculation 
for the core nucleus. 
We neglect the recoil term in Eq. (\ref{eq:3bWS}) because the core ($A_C=100$) is much heavier than nucleons. 
For the residual interaction $V_{\rm res}$, we take a density-dependent zero-range interaction 
which has both spin-singlet and spin-triplet channels, 
\begin{eqnarray}
V_{\rm res}(\bvec r_p,\bvec r_n)&=&
\hat P_s V_0\delta(\bvec r_p-\bvec r_n)\left[1+x_s\left(\frac{\rho(r_p)}{\rho_0}\right)^{\alpha_s}\right]
\nonumber \\
&&
+\hat P_t fV_0\delta(\bvec r_p-\bvec r_n)\left[1+x_t\left(\frac{\rho(r_p)}{\rho_0}\right)^{\alpha_t}\right], 
\nonumber \\
\label{eq:vpair}
\end{eqnarray}
where $\hat P_s$ and $P_t$ are the projectors onto spin-singlet and spin-triplet channels, respectively, 
\begin{eqnarray}
\hat P_s&=&\frac{1}{4}-\frac{1}{4}\bvec\sigma(p)\cdot\bvec\sigma(n)\\
\hat P_t&=&\frac{3}{4}+\frac{1}{4}\bvec\sigma(p)\cdot\bvec\sigma(n). 
\end{eqnarray}
Note that, when two nucleons are in the same $j$ shell, the spin-singlet interaction 
acts only on even-$J$ states [$J^\pi=0^+, 2^+,...,(2j-1)^+$], while the spin-triplet interaction acts on 
odd-$J$ [$J^\pi=1^+, 3^+,..., (2j)^+$].  
The density $\rho(r)$ in the density dependent term in Eq. (\ref{eq:vpair}) is given by the 
mean-field calculation for the core. 
In order to  fix the $S=0$ coupling strength of the delta interaction $V_0$, 
we first introduce the cutoff energy $E_{\rm cut}$ of the model space and 
adjust $V_0$ so that the empirical neutron pairing gap of $^{101}$Sn is reproduced. 
The ratio $f$ of the strength of the spin-triplet interaction to that of the singlet one is estimated using the data 
of other $N=Z$ odd-odd nuclei. 
See Sec. \ref{sec:result} for details how to fix the parameters.

The Hamiltonian is diagonalized on two-particle basis of the valence nucleons, 
which is given by a product of proton and neutron single-particle states. 
The single-particle states are the eigenstates of the self-consistent mean-field Hamiltonian, 
\begin{eqnarray}
h_{\rm HF}(\tau)|\varphi_{nljm}^{(\tau)}\rangle = \epsilon_{nlj}^{(\tau)}|\varphi_{nljm}^{(\tau)}\rangle
\ (\tau = p\ {\rm or}\ n),
\end{eqnarray}
where $n$, $l$, $j$, and $m$ are the radial quantum number, the orbital angular momentum, the total 
angular momentum, and its projection on $z$-axis, respectively. 
$\tau=p\ {\rm or}\ n$ labels the isospin of the nucleon. 
The two-particle basis coupled to a good angular momentum $J,M$ is constructed as 
\begin{eqnarray}
|ab,JM\rangle = \bigl[|\varphi^{(p)}_{n_al_aj_a}\rangle\otimes|\varphi^{(n)}_{n_bl_bj_b}\rangle\bigr]_{JM}. 
\end{eqnarray}
Notice that we take the proton-neutron formalism and do not antisymmetrize the state, in order to 
take into account the breaking of isospin symmetry due to the Coulomb interaction. 
The single-particle states below $E_{\rm cut}$ are included in the model space. 
The continuum states are discretized within a spherical radial box. 

We use the model described above 
to get solutions for the unbound $^{102}$Sb nucleus, 
where the valence proton orbitals are all in continuum. 
Although the low-lying states of the system should be resonances,  
we do not treat explicitly the continuum  in our calculations. 
Instead, we interpret our model in the following way. 
The single-particle state is either resonance-like (large amplitude inside the core of the nucleus) or 
non-resonant state when they are confined within a  large radial box \cite{stab1,stab2}. 
We simply regard a solution obtained by the diagonalization as a ``three-body resonance'' 
if it is dominated by resonance-like single-particle states. 
In this way we get approximate resonance energies and the structure of the states, but not the resonance widths.

\subsection{Relativistic version of the model}
We also perform the three-body model calculations in the relativistic framework. 
That is, we construct the model space given by a self-consistent calculation with a relativistic energy density functional. 
Since the nucleons are described as Dirac spinors, we need to modify some of the operators as well. 
In the relativistic theory, the spin operator of a Dirac particle is given as \cite{Sakurai}
\begin{eqnarray}
\bvec S=\frac{\bvec\Sigma}{2}=
\left(\begin{array}{cc}
\frac{\bvec\sigma}{2} & 0 \\
0 & \frac{\bvec\sigma}{2}
\end{array}\right), 
\end{eqnarray}
which obeys the SU(2) algebra, $[S_i,S_j]=i\epsilon_{ijk}S_k$. 
We give the relativistic version of the projectors by replacing $\bvec\sigma$ by $\bvec\Sigma$, 
\begin{eqnarray}
\hat P_s=\frac{1}{4}-\frac{1}{4}\bvec\Sigma(p)\cdot\bvec\Sigma(n),\ 
\hat P_t=\frac{3}{4}+\frac{1}{4}\bvec\Sigma(p)\cdot\bvec\Sigma(n), 
\end{eqnarray}
which project a two-particle state onto the spin-singlet or spin-triplet representation.  
The basis states are constructed using the 
four-component Dirac wave functions of proton and neutron in a similar way to the non-relativistic model.  
See Appendix \ref{app:3brel} for the details of the formalism.

\section{Choice of energy density functional}\label{sec:edf}

\begin{figure}
\begin{center}
\includegraphics[scale= .68,angle= 0]{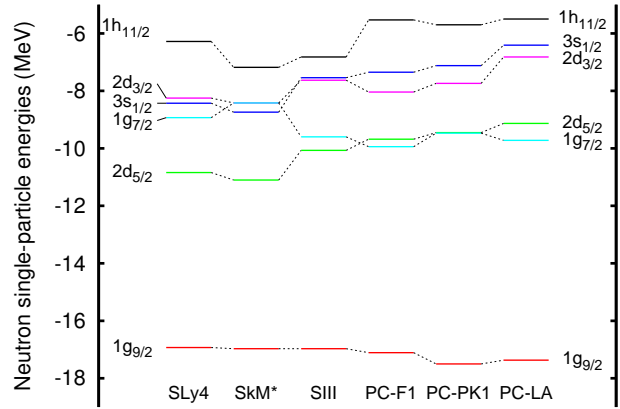}
\end{center}
\caption{(Color online) Neutron single-particle energies in $^{100}$Sn nucleus with SkM$^*$, SIII, SLy4, PC-F1, PC-PK1, 
PC-LA functionals. }
\label{fig:esp_rel}
\end{figure}

An important ingredient of the present model is the choice of the energy density functional (EDF) 
used for the mean-field part. 
The configuration of the nucleons in the valence space is quite sensitive to the single-particle energies. 
Since the structure of single-particle levels are substantially different among EDFs, the result of the three-body model calculation 
may depend sensitively on the EDF. 
In order to illustrate the variation of single-particle energies, we show in Fig. \ref{fig:esp_rel} 
a comparison of the neutron single-particle energies in $^{100}$Sn nucleus 
among three non-relativistic (SLy4 \cite{sly4}, SkM$^*$ \cite{skms}, and SIII \cite{s3}) 
and three relativistic point-coupling  (PC-F1 \cite{Bur02}, PC-PK1 \cite{PCPK1}, and PC-LA \cite{PCLA}) EDFs. 
Although the Fermi energies are similar, the level spacings and orderings differ among them.

\begin{figure}
\begin{center}
\includegraphics[scale= .68,angle= 0]{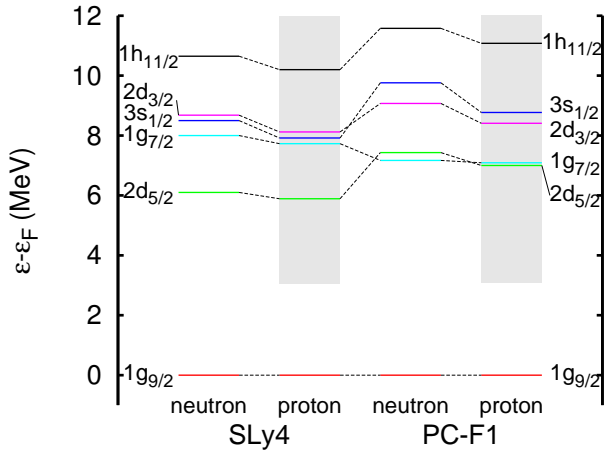}
\end{center}
\caption{(Color online) Neutron and proton single-particle energies with respect to the Fermi energies in $^{100}$Sn nucleus with 
SLy4 and PC-F1. The shaded area indicates the continuum-energy region. 
All the proton levels shown here are resonances except for the the Fermi level ($1g_{9/2}$). }
\label{fig:esp_pn}
\end{figure}

We see in particular that, with the relativistic interactions, the $1g_{7/2}$ level is nearly degenerate 
to the $2d_{5/2}$ level. 
In contrast, with the Skyrme functionals except for SIII, $1g_{7/2}$ level is located well above the 
$2d_{5/2}$ state and close to the $2d_{3/2}$ level. 
This difference may induce  a substantial effect on the structure of the system and $(T=0, S=1)$ 
pair coupling
because it prevents the valence nucleons from efficiently gaining the pairing energy in $(d)^2$ configurations. 
This near degeneracy with the relativistic functionals is due to the pseudo-spin symmetry \cite{Gino05,Liang15}, 
which originates from the cancellation of large scalar and vector fields. 
Thus it is a commonly observed feature of the relativistic mean-field theories for nuclei. 
Note that a $\gamma$ ray of 172 keV between the first-excited and the ground states of 
$^{101}$Sn has been measured in Refs. \cite{Se07,Da10}, which is interpreted as a single-neutron transition 
between $d_{5/2}$ and $g_{7/2}$ levels. 
In Ref. \cite{Se07}, a shell-model analysis on the systematic trend of low-lying spectra in Sn isotopes 
has been made to assign $J^\pi = 5/2^+$ and $7/2^+$ for the ground and the first excited states of $^{101}$Sn, 
respectively. 
On the other hand, the $\alpha$-decay branching ratios measured in Ref. \cite{Da10} for a chain 
${}^{109}{\rm Xe}\to{}^{105}{\rm Te}\to{}^{101}{\rm Sn}$ suggest the opposite level ordering in $^{101}$Sn. 
The shell-model analyses performed in Ref. \cite{Da10} also support their assignment. 
According to the discussions in the literature \cite{Se07,Da10,Li06,Lo12}, the ordering of neutron $d_{5/2}$ and $g_{7/2}$ states in $^{101}$Sn is still controversial and 
not yet clearly determined. 
However, we emphasize that the near degeneracy of these two single-particle states can be well reproduced by the 
relativistic mean field. 
It is worth mentioning that the calculated single-particle proton states  have the similar feature to that of the neutron states. 
In Fig. \ref{fig:esp_pn}, we show both proton and neutron single-particle energies in the $^{100}$Sn 
core  
nucleus obtained with SLy4 and PC-F1 functionals. 
Note that all the proton levels above the Fermi level are in the continuum, which is indicated 
with the shaded areas. 
The energies of the proton levels shown here are the resonance energies obtained with the stabilization 
method \cite{stab1,stab2}. 
As expected, 
we see that the structure of single-particle energies are similar between the neutron and proton states, and 
that $1g_{7/2}$ and $2d_{5/2}$ proton states  also are almost degenerate for the relativistic functional.

\begin{widetext}

In order to see a more systematic aspect of EDFs, in Table \ref{tb:esp} 
we summarize the comparison of single-particle energies of the lowest-unoccupied 
orbital in $N=Z$ doubly-magic systems for the relativistic and non-relativistic EDFs. 
Experimental values of the neutron (proton) separation energies of 
the doubly-magic $+n$ ($+p$) 
nuclei are also shown in the table.  
We see that the Skyrme interactions systematically underestimate the single-particle energies 
for lighter systems, $A=16$, 40, and 56. 
On the other hand, relativistic functionals are in better agreement with the data for all the $A=16$, 40, and 
for some cases in $A=56$ nuclei. 
For $A=100$, SIII and SLy4 appear to be better than the relativistic functionals. 
We should note also that not only the lowest-unoccupied level ($j_>$) but also the spin-orbit splitting with 
its partner ($j_<$) is also quite important for the $T=0$ pair coupling  since $(j_>)^2$ and $(j_>j_<)$ configurations 
coherently contribute to render the wave functions more $LS$-coupling-like and gain $T=0$ pairing energy 
\cite{Sa13}. 

\begin{table}
\caption{
 Single-particle energies of the lowest-unoccupied orbitals in $N=Z$ doubly-closed systems. 
The neutron (proton) separation energies of the doubly-magic $+n$ ($+p$) nuclei 
are shown together as experimental values. 
All the energies are shown in unit of MeV. Note that the lowest-unoccupied neutron orbital in $^{100}$Sn are $2d_{5/2}$ for the Skyrme models while those for 
PC-F1, PC-PK1, and PC-LA are  $1g_{7/2}$. 
 }
\begin{center}
\begin{tabular}{cc|c|ccc|ccc}
\hline\hline
&               & expt.  & SLy4 & SkM$^*$ & SIII & PC-F1 & PC-PK1 & PC-LA \\
\hline
$^{16}$O  & p & $-0.6$ & $-3.5$  & $-4.1$ & $-3.6$ & $-1.5$ & $-1.6$ & $-1.5$\\
 $1d_{5/2}$ & n & $-4.1$ & $-6.7$ & $-7.3$ & $-6.9$ & $-4.9$ & $-5.1$ & $-5.0$ \\
\hline
$^{40}$Ca & p & $-1.1$  & $-2.8$& $-3.3$  & $-3.1$ & $-1.5$ & $-1.6$ & $-1.8$\\
 $1f_{7/2}$ & n & $-8.4$  & $-9.6$& $-10.1$  & $-9.9$ & $-8.5$ & $-8.7$ & $-8.8$\\
\hline
$^{56}$Ni  & p & $-0.7$ & $-2.1$& $-2.3$  & $-1.2$ & $-0.8$ & $-0.1$ & $-0.3$ \\
 $2p_{3/2}$ & n & $-10.2$  & $-11.1$ & $-11.3$ & $-10.2$ & $-9.8$ & $-9.2$ & $-9.3$\\
\hline
$^{100}$Sn  & p & --   & --  & -- & -- & -- & -- & -- \\
 $2d_{5/2} (1g_{7/2}) $ & n & $-11.1$  & $-10.8$& $-7.6$ & $-10.1$ & $(-9.9)$ & $(-9.5)$ & $(-9.7)$\\
\hline\hline
\end{tabular}
\end{center}
\label{tb:esp}
\end{table}

\end{widetext}

To summarize this section, in particular for $A=100$, relativistic functionals reproduce well 
the empirical single-particle energies such as  the near degeneracy of $1g_{7/2}$ and $2d_{5/2}$ levels. 
The Skyrme interaction SIII has a  similar feature to the relativistic models in this respect. 
Regarding the systematics of single-particle energies in $N=Z$ closed-shells nuclei besides  $A=100$, 
the relativistic functionals give  better results for lighter systems while some of the Skyrme functionals 
are better for heavier systems. 
In the present study, we focus on the effect of the $1g_{7/2}-2d_{5/2}$ degeneracy 
on the structure of $^{102}$Sb system.  
We use the six EDFs illustrated here in our three-body model calculations and compare the results.

\section{Three-body model calculations for $^{102}$Sb}\label{sec:result}

\subsection{Determination of parameters and numerical setup}
Now we perform the three-body model calculations.  
For the mean-field part, we use the six non-relativistic and relativistic interactions illustrated in the last section. 
For the residual interaction, we take a surface type ($x_s=x_t=-1$, $\alpha_s=\alpha_t=1$, 
and $\rho_0=0.16$ fm$^{-3}$) and $E_{\rm cut}=20$ MeV for all EDFs. 
Note that this cutoff is large enough to cover the valence major shells of neutrons and protons, 
which are expected to give dominant contributions to the low-lying states. 
The radial coordinate is discretized up to $R_{\rm box}=40$ fm with the mesh size of $\Delta r =0.1$ fm. 
In the model space, we take the single-particle states with the orbital angular momentum up to $l_{\rm max}=7$. 
We have checked the convergence of the results with respect to $l_{\rm max}$ and 
also the stability of the results against $R_{\rm box}$. 

For the given $E_{\rm cut}$, the strength $V_0$ is adjusted to reproduce the 
empirical pairing gap in $^{101}$Sn,  
\begin{eqnarray}
\Delta=\frac{1}{2}\left[S_{2n}(^{102}{\rm Sn})-2S_n(^{101}{\rm Sn}) \right]= 0.8\ {\rm MeV}.  
\end{eqnarray}
The data is taken from Ref. \cite{ame12}. 
The gap is evaluated in our model by a formula 
\begin{eqnarray}
\Delta = \frac{1}{2}\left[-E(^{102}{\rm Sn})-2(-\epsilon^{(n)}_{LU})\right].   
\label{eq:gap}
\end{eqnarray}
Here $E(^{102}{\rm Sn})$ is energy of the $0^+$ ground state of 
$^{102}{\rm Sn}={^{100}{\rm Sn}}+2n$. 
This is obtained within our model by replacing the basis by two-neutron states. 
$\epsilon^{(n)}_{LU}$ is the neutron single-particle energy of the lowest unoccupied state 
in $^{100}$Sn for a given EDF. 
The values of $V_0$ fixed for $E_{\rm cut}=20$ MeV are summarized in Table \ref{tb:vs}. 

The ratio of the spin-triplet to spin-singlet interactions $f$ is yet to be determined. 
In order to estimate its value, we use the data of known $N=Z$ doubly-magic plus two nucleons 
systems with $A=16+2$, $40+2$, and $56+2$. 
For each $A$, we first determine $V_0$ in the same manner as for $A=102$, then adjust $f$ so that 
the measured energy difference $E(1^+_1)-E(0^+_1)$ in the odd-odd (${\rm core}+p+n$) system 
is reproduced. (See Fig. \ref{fig:spec} in Sec. \ref{ssec:result} for the experimental data.) 
The values of the ratio $f$ so determined are tabulated in Table \ref{tb:f}. 
Interestingly, the resultant values are all larger than unity, and 
they are almost constant for a given system with different EDFs used for the mean-field. 
It is also found that $f$ ranges from $1.2$ to $1.5$, which makes it difficult to determine a unique value for $f$. 
Therefore, for our calculations of $A=102$, we vary the value of $f$ within this range: $f=1.2$, 1.3, 1.4, and 1.5. 
Note that similar values, $f\sim 1.1$ for $^{42}$Sc and $f\sim 1.4$ for $^{58}$Cu, are obtained 
in Ref.  \cite{Yo14} with a particle-particle random-phase approximation approach, 
while  a value $f\sim 1.6$ for $sd-$ and $pf-$shells is extracted from shell model Hamiltonians in Ref. \cite{BeLu10}. 

\begin{table}
\caption{The spin-singlet interaction strength $V_0$ for $E_{\rm cut}=20$ MeV. }
\begin{center}
\begin{tabular}{ccccccc}
\hline\hline
Force & SLy4 & SkM$^*$  & SIII   & PC-F1 & PC-PK1 & PC-LA \\
$V_0$ (MeV fm$^3$) & $-731$ &$-754$ &$-623$ &$-602$ &$-555$ &$-675$ \\
\hline\hline
\end{tabular}
\end{center}
\label{tb:vs}
\end{table}

\begin{table}
\caption{The ratio $f$ of the spin-triplet to -singlet interaction strength fitted to the data of 
$E(1^+_1)-E(0^+_1)$ in $^{18}$F, $^{42}$Sc, $^{58}$Cu nuclei for each EDF. 
The cutoff energy is taken to be $E_{\rm cut}=20$ MeV. }
\begin{center}
\begin{tabular}{ccccccc}
\hline\hline
      & SLy4 & SkM$^*$ & SIII & PC-F1 & PC-PK1 & PC-LA \\
\hline
  $^{18}$F & 1.48 & 1.50 & 1.51 & 1.43 & 1.45 & 1.52 \\
$^{42}$Sc & 1.18 & 1.20 & 1.20 & 1.15 & 1.17 & 1.20 \\
$^{58}$Cu & 1.31 & 1.30 & 1.33 & 1.29 & 1.28 & 1.34 \\ 
\hline\hline
\end{tabular}
\end{center}
\label{tb:f}
\end{table}

\subsection{Results and discussion}\label{ssec:result}

\subsubsection{Low-lying energy spectrum and Gamow-Teller-type transitions}

\begin{figure}
\begin{center}
\includegraphics[scale= .65,angle=0 ]{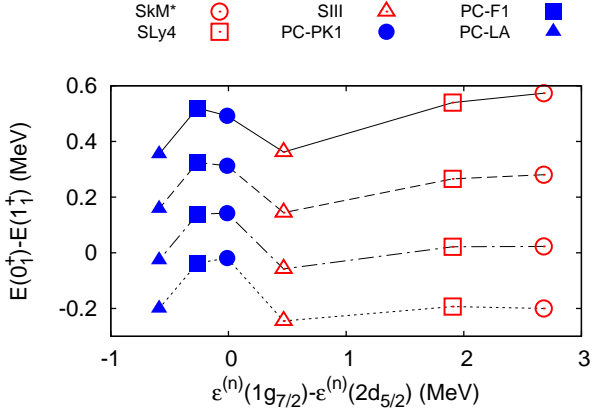}
\end{center}
\caption{(Color online) The excitation energy from $1^+_1$ to $0^+$ state 
in $^{102}$Sb as a function of $\epsilon^{(n)}_{1g_{7/2}}-\epsilon^{(n)}_{2d_{5/2}}$. 
The symbols connected by dotted, dot-dashed, dashed, and solid lines are given with 
$f=1.2$, 1.3, 1.4, and 1.5, respectively. }
\label{fig:exc-de}
\end{figure}

First, we show in Fig. \ref{fig:exc-de} 
the energy difference between $0^+_1$ and $1^+_1$ levels in $^{102}$Sb, $E(0^+_1)-E(1^+_1)$,
obtained with different values of $f$ and different EDFs. 
In the figure, the horizontal axis is taken to be the 
difference of single-particle energies between the neutron pseudo-spin-orbit partners, 
$\Delta\epsilon=\epsilon^{(n)}_{1g_{7/2}}-\epsilon^{(n)}_{2g_{5/2}}$, 
and the symbols connected with the dotted, dot-dashed, dashed, and solid lines are obtained  with 
$f=1.2$, 1.3, 1.4, and 1.5, respectively. 
The energies of $1^+_1$ and $0^+_1$ levels in $N=Z$ doubly-magic core + p + n nuclei 
would be a good playground  of competition between $(T=0,S=1)$ and $(T=1,S=0)$ pair couplings \cite{Tani14}. 
It is found that the $0^+_1-1^+_1$ energy difference does not have a strong correlation with $\Delta\epsilon$, 
and that $f\gtrsim 1.3$ makes the $1^+$ state be the ground state. 

\begin{figure}
\begin{center}
\includegraphics[scale= .45,angle=0 ]{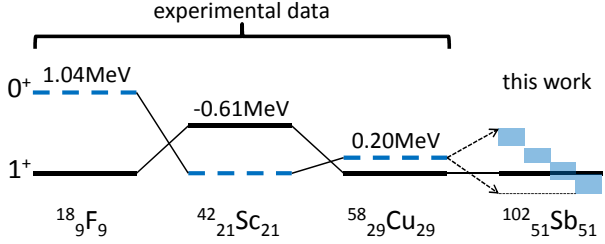}
\end{center}
\caption{(Color online) $J^\pi=1^+_1$ and $J^\pi=0^+_1$ energies for $N=Z$ odd-odd nuclei. 
For $^{18}$F, $^{42}$Sc, and $^{58}$Cu, measured energy differences $E(0^+_1)-E(1^+_1)$ are shown. 
For $^{102}$Sb the result of our calculation is shown. Four blue boxes indicate the $0^+_1$ energy 
obtained with $f=1.5$, 1.4, 1.3, 1.2 from the left to right, 
with respect to the $1^+_1$ energy which is shown by black solid line. 
The vertical width of each box corresponds to the range of deviation over the six EDFs used for the 
mean-field. }
\label{fig:spec}
\end{figure}

Figure \ref{fig:spec} shows  a comparison of measured $1^+_1$ and $0^+_1$ energies 
in other $N=Z$ odd-odd systems and that calculated  in the present study  for $^{102}$Sb. 
The shaded boxes for $^{102}$Sb indicate the $0^+_1$ energy relative to the $1^+_1$ energy 
with $f=1.5$, 1.4, 1.3, and 1.2, from left to right. The vertical width of each shaded box represents the range of deviation 
among the different EDFs for the given $f$. 
In the experimental data for the lighter systems, 
the inversion of the $1^+_1$ and $0^+_1$ energies in $^{42}$Sc could be interpreted as 
the effect of larger spin-orbit splitting in $1f$ orbitals occupied by the valence nucleons 
\cite{Tani14}, whereas it is $1d$ and $2p$ orbitals  that are occupied in $^{18}$F and $^{58}$Cu, respectively. 
In $^{102}$Sb where $2d$ orbitals are available in the valence space similar to $1d$ orbitals  in $^{18}$F, 
both $1^+_1$ and $0^+_1$ can be the ground state within the range of $f$ taken in this work.

\begin{figure}
\begin{center}
\includegraphics[scale= .65,angle=0 ]{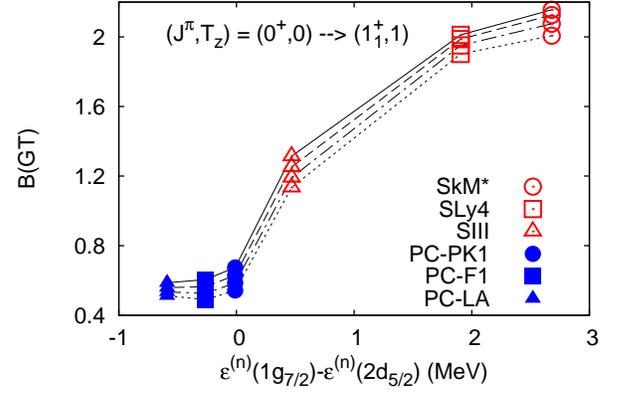}
\end{center}
\caption{(Color online) $B(GT)$ values for the transition from the ground state of $^{102}$Sn to the first $1^+$ state of 
$^{102}$Sb as a function of $\epsilon^{(n)}_{1g_{7/2}}-\epsilon^{(n)}_{2d_{5/2}}$. 
The symbols connected by dotted, dot-dashed, dashed, and solid lines are given with 
$f=1.2$, 1.3, 1.4, and 1.5, respectively.}
\label{fig:bgt-de}
\end{figure}

In Fig. \ref{fig:bgt-de}, we plot  the  Gamow-Teller (GT) transition probability 
\begin{eqnarray}
B(GT) &=& |\langle^{102}{\rm Sb}(1^+_1)\|\bvec O(GT_-)\|^{102}{\rm Sn}(0^+_1)\rangle|^2,
\end{eqnarray}
as a function of the energy difference $\Delta\epsilon=\epsilon^{(n)}_{1g_{7/2}}-\epsilon^{(n)}_{2g_{5/2}}$.  
The GT transition operator  is taken to be 
\begin{eqnarray}
\bvec O(GT_{\pm}) = \sum_i\bvec \sigma(i)t_\pm(i), 
\end{eqnarray}
for the non-relativistic cases and 
\begin{eqnarray}
\bvec O(GT_{\pm}) = \sum_i\bvec\Sigma(i)t_\pm(i), 
\end{eqnarray}
for the relativistic ones, where $i$ runs over the two valence nucleons in our model. 
One sees that the $B(GT)$ is remarkably sensitive to $\Delta\epsilon$, 
or EDF used for the mean field, as compared to the $0^+_1-1^+_1$ energy difference. 
Its sensitivity is much larger than that against the variation of $f$ value. 

\begin{figure}
\begin{center}
\includegraphics[scale= .65,angle=0 ]{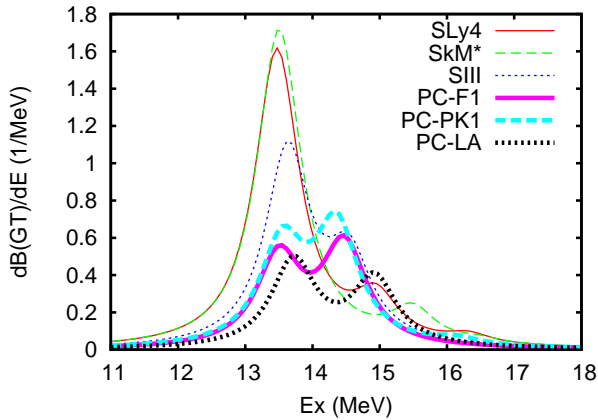}
\end{center}
\caption{(Color online) Gamow-Teller strength distributions as a function of the excitation energy from the ground state of $^{102}$Sn. 
The discrete peaks obtained from the model are smeared by Lorentzian with 0.4 MeV width. }
\label{fig:bgt-ex}
\end{figure}

\begin{figure}
\begin{center}
\includegraphics[scale= .8,angle=0 ]{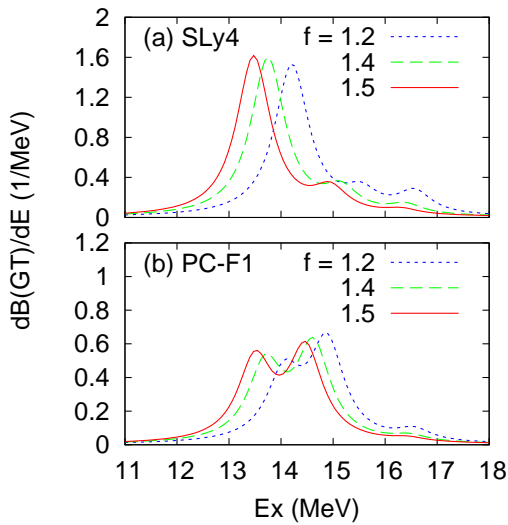}
\end{center}
\caption{(Color online) Gamow-Teller strength distributions obtained with (a) SLy4 and (b) PC-F1 
for three different values of the triplet-to-singlet ratio $f$. 
As in Fig. \ref{fig:bgt-ex}, the original $B(GT)$ is smeared by Lorentzian with 0.4 MeV width. }
\label{fig:bgt-f}
\end{figure}

We also show in Fig. \ref{fig:bgt-ex} the $B(GT)$ distribution as a function of the excitation energy 
with respect to the $0^+_1$ state of $^{102}$Sn. 
The clear difference between non-relativistic and relativistic cases are observed; 
the strength distribution is concentrated into the first peak for SLy4 and SkM* 
whereas it is fragmented into the first and second peaks for PC-F1,  PC-PK1, and PC-LA. 
SIII that has smaller $\Delta\epsilon$ shows a feature in between them. 
Figure \ref{fig:bgt-f} shows the $f$-dependence of GT distribution within the range of $f=(1.2\sim1.5)$.   
We compare the distributions obtained with SLy4 (upper panel) 
and with PC-F1 (lower panel) for the three different values of $f=1.2$, 1.4, and 1.5. 
One sees that the characteristic difference between them remains unchanged for different values of $f$. 
For both SLy4 and PC-F1, the peaks are shifted toward lower energy as $f$ increases. 
At the same time, the strength of the first peak absorbs some strength from the higher states. 
Similar $f$-dependence of $B(GT)$ distributions have also been observed for other $N=Z$ nuclei with 
Skyrme Hartee-Fock-Bogoliubov + quasi-particle random-phase approximation \cite{Bai13}. 

Here let us remark the sum rule of the GT transition. 
Gamow-Teller $\beta^+$ decays of $^{102}$Sn have been measured 
in Refs. \cite{Faester02} and \cite{Karny06}, where the sum of the $B(GT_+)$ values of observed transitions 
are deduced to be $4.0(6)$ and $4.2(8)$, respectively. 
If we take into account the $B(GT)$ deduced in Refs. \cite{Faester02,Karny06}, the sum rule tells us that 
\begin{eqnarray}
\sum_f B(GT_-,^{102}{\rm Sn}\to f) \gtrsim 10. 
\end{eqnarray}
since the GT Ikeda sum rule is 
\begin{eqnarray}
&&\sum_{f}|\langle f|\bvec O(GT_-)|i\rangle|^2
-\sum_{f'}|\langle f'|\bvec O(GT_+)|i\rangle|^2 \nonumber \\
&=& 3(N-Z)=6    
\end{eqnarray}
for $^{102}{\rm Sn}$.  
This means that the low-lying GT states obtained in the present calculations only exhausts 
about 10 \% for the relativistic models  and 20\% for the non-relativistic models of the total GT$_-$ strength. 
The large part of the sum lies in higher excitation energy. 
This is due to the fact  that there is highly collective giant GT resonance involving the nucleon 
excitations from $g_{9/2}$ state to $g_{7/2}$ state, which is beyond our model space.

In order to understand the large difference of $B(GT)$ between non-relativistic and relativistic models, 
in Table \ref{tb:wf} we summarize the information of the wave functions obtained with SLy4 and PC-F1. 
In the table we compare the composition of the wave functions and the probability of spin triplet $P(S=1)$ 
in the $0^+$ ground state of $^{102}$Sn and the low-lying states of $^{102}$Sb. 
With SLy4, both $0^+_1$ and $1^+_1$ states are dominated by $(d)^2$ configurations.  
Furthermore, the $1^+_1$ state has the large probability of spin triplet $P(S=1)=81.5$ \% 
while the $0^+_1$ has $P(S=0)=1-P(S=1)=75.9$ \%. 
This implies that they belong to the same SU(4) multiplet 
in a relatively  good approximation as was observed also in $^{18}$F \cite{Tani14}. 
Thus this state has the similar nature as the ``low-energy super GT'' state observed in $^{42}$Sc in Ref. \cite{Fu14}. 
This explains the large $B(GT)$ values between the two states obtained with the nonrelativistic EDFs. 

On the other hand, $0^+_1$ and $1^+_1$ states with PC-F1 is dominated by $(g_{7/2})^2$ or 
$d_{5/2}\times g_{7/2}$ configurations since the $g_{7/2}$ orbital is almost degenerated  with  the 
$d_{5/2}$ orbital  at nearly the same energy. 
Therefore, despite that the energy difference between $0^+_1$ and $1^+_1$ states is similar to that of 
the non-relativistic models, 
the wave function of the first $1^+$ state with PC-F1 is far different from the $(L,S,T)=(0,1,0)$ coupling scheme. 
The transition from $0^+_1$ to $1^+_1$ is made mainly by promoting a nucleon from 
$g_{7/2}$ to $d_{5/2}$, which is forbidden in GT-type transitions. 
Although it is allowed relativistically by the transition between the lower components of Dirac spinor \cite{Gino05}, 
its contribution is only $\sim 1$ \% of the total transition amplitude. 
This is why the $B(GT)$ to $1^+_1$ is weaker in the relativistic models.

\begin{table}
\caption{
The occupation probabilities of each configuration in \%
obtained with $f=1.5$ for SLy4 and PC-F1. 
The configuration which has the largest probability in each state is shown with bold-face letters. 
In the last row are shown the probabilities in \% of spin triplet component in the wave function. 
}
\begin{center}
\begin{tabular}{c|cc|c|cc|c}
\hline\hline
        & \multicolumn{3}{c}{SLy4} &  \multicolumn{3}{|c}{PC-F1}\\
\hline
        & \multicolumn{2}{c|}{$^{102}$Sb} &$^{102}$Sn
        & \multicolumn{2}{c|}{$^{102}$Sb} &$^{102}$Sn \\
        & $~~1^+_1~~$ &  $~~0^+_1~~$ & $~0^+_1~$ 
& $~~1^+_1~~$ & $~~~0^+_1~~$ & $~0^+_1~$ \\
\hline
$(d_{5/2})^2$                      & {\bf 59.3} & {\bf 84.1} & {\bf 78.7}        & 17.0 & 32.9 & 26.8 \\
$(d_{5/2}\times d_{3/2})$	& 13.9  & 0 & 0                                               & 2.3 & 0 & 0\\
$(g_{7/2})^2$                     & 6.5 & 7.0 & 8.7                             & 32.4 & {\bf 59.4} & {\bf 64.3} \\
$(d_{5/2}\times g_{7/2})$  & 5.7  & 0 & 0                                        & {\bf 38.9}  & 0 & 0 \\
\hline
$P(S=1)$                         & 81.5 & 26.0 & 24.1                                     & 82.7  & 42.2 & 43.4\\
\hline\hline
\end{tabular}
\end{center}
\label{tb:wf}
\end{table}

\subsubsection{Systematics of the other physical quantities}

\begin{figure}
\begin{center}
\includegraphics[scale= .65,angle=0 ]{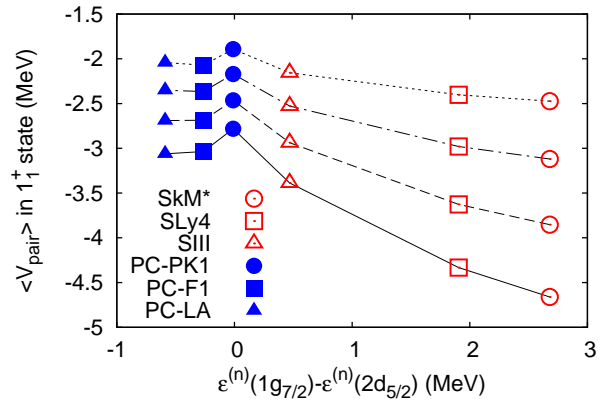}
\end{center}
\caption{(Color online) The expectation value of the residual interaction $\langle V_{\rm res}\rangle$ 
in the first $1^+$ state of 
$^{102}$Sb as a function of $\epsilon^{(n)}_{1g_{7/2}}-\epsilon^{(n)}_{2d_{5/2}}$. 
The symbols connected by dotted, dot-dashed, dashed, and solid lines are given with 
$f=1.2$, 1.3, 1.4, and 1.5, respectively. }
\label{fig:ep-de}
\end{figure}

\begin{figure}
\begin{center}
\includegraphics[scale= .65,angle=0 ]{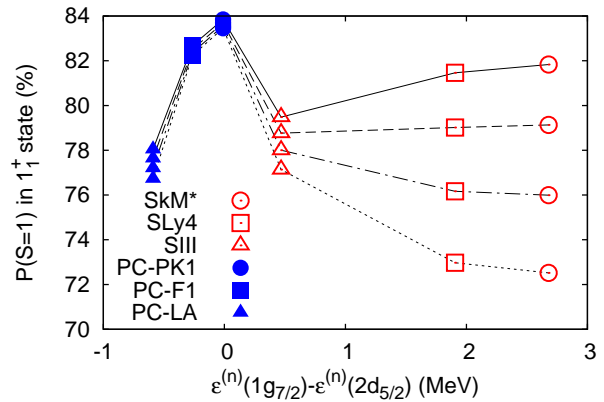}
\end{center}
\caption{(Color online) Probability of spin triplet $P(S=1)$ in the first $1^+$ state of 
$^{102}$Sb as a function of $\epsilon^{(n)}_{1g_{7/2}}-\epsilon^{(n)}_{2d_{5/2}}$. 
The symbols connected by dotted, dot-dashed, dashed, and solid lines are given with 
$f=1.2$, 1.3, 1.4, and 1.5, respectively.}
\label{fig:pt-de}
\end{figure}

\begin{figure}
\begin{center}
\includegraphics[scale= .65,angle=0 ]{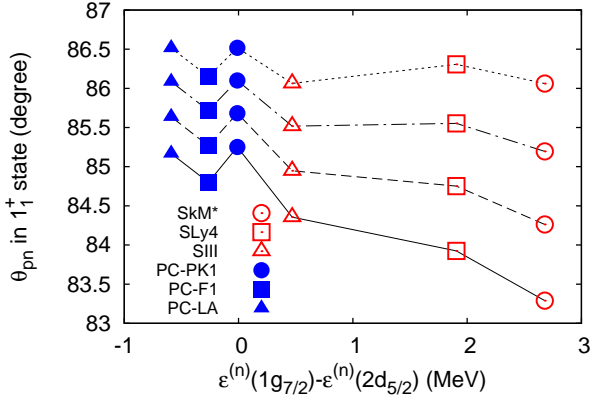}
\end{center}
\caption{(Color online) Opening angle $\theta_{pn}$ between the valence nucleons in the first $1^+$ state of 
$^{102}$Sb as a function of $\epsilon^{(n)}_{1g_{7/2}}-\epsilon^{(n)}_{2d_{5/2}}$. 
The symbols connected by dotted, dot-dashed, dashed, and solid lines are given with 
$f=1.2$, 1.3, 1.4, and 1.5, respectively.}
\label{fig:ang-de}
\end{figure}

Here we discuss the systematics of other quantities. 
We show in Figs. \ref{fig:ep-de}, \ref{fig:pt-de}, and \ref{fig:ang-de} the systematics of 
the expectation value of the residual interaction, $\langle 1^+_1 |V_{\rm res}|1^+_1 \rangle$, 
the probability of spin triplet, $P(S=1)$,  and 
the opening angle between valence nucleons, 
\begin{eqnarray}
\theta_{pn} = \cos^{-1}\left(\left\langle 1^+_1\left|\frac{\bvec r_p\cdot\bvec r_n}{r_pr_n}
\right|1^+_1\right\rangle\right), 
\end{eqnarray}
respectively, in the $1^+_1$ state. 

The value of $\langle V_{\rm res} \rangle$ shown in Fig. \ref{fig:ep-de} exhibits a strong correlation 
with $\Delta\epsilon$ as $f$ increases, 
which directly shows that a larger  occupation of the $g_{7/2}$ orbit hinders the $S=1$ 
coupling in $(d)^2$ configurations. 
Another observation is that
the larger is $\Delta\epsilon$, the more sensitive to $f$ are 
the values of $\langle V_{\rm res} \rangle$ and also $P(S=1)$ (Fig. \ref{fig:pt-de}). 
This shows that the the coupling of a spin-triplet pair 
involving the $(d_{5/2})^2$ and $d_{5/2}\times d_{3/2}$ 
configurations is sensitive to the strength of the $S=1$ interaction as expected.  
By contrast, the coupling built with $(d_{5/2})^2$ and $d_{5/2}\times g_{7/2}$ configurations is less sensitive which was also expected. 

The opening angle $\theta_{pn}$ between valence nucleons shown in Fig. \ref{fig:ang-de} 
is a measure of the deuteron-like spatial localization of the valence nucleons, which is 
analogous to the di-neutron pair in neutron-rich systems. 
We see that the angle is little less than $90^\circ$ for any case examined in the present work. 
Thus, as long as the mean value of the angle is concerned, the deuteron-like cluster is not likely to exist in $^{102}$Sb. 
It is an expected consequence because the valence neutron would be deeply bound in the shell-model orbital 
in this neutron-deficient system, in contrast to the di-neutron in neutron-rich nuclei for which one may draw 
a picture of the di-neutron in the same orbit loosely coupled to the core. 
The angle of deuteron-like pair may be compared with the angle of di-neutron configuration in the other theoretical works for neutron-rich nuclei, 
$67.9^\circ$ for $^6$He in Ref. \cite{Kiku10} and $66.33^\circ$, $65.29^\circ$, and $82.37^\circ$ 
for $^6$He, $^{11}$Li, and $^{24}$O, respectively in Ref. \cite{Hagi05}.

\section{Summary and perspectives}\label{sec:summary}
We have applied a new three-body model based on self-consistent mean-field calculations, 
which can describe the coexistence of $T=1$ and $T=0$ coupling effects in the low-lying levels, 
to $^{102}{\rm Sb} = {}^{100}{\rm Sn}+p+n$ nucleus. 
The low-lying states in the nucleus 
were discussed in terms of the relation between its spectroscopic properties and the 
coupling between the two valence nucleons in $T=0$ channel. 
The $T=1$ interaction strength is fixed by the empirical neutron pairing gap, 
and the $T=0$ strength is estimated from the $1^+_1$ and $0^+_1$ energy differences in 
known $N=Z$ odd-odd nuclei. 

We have adopted  three non-relativistic and three relativistic EDFs which have 
different features in the  valence single-particle levels. 
Experimental data \cite{Se07,Da10} for $^{101}$Sn support the near degeneracy of $1g_{7/2}$ and $2d_{5/2}$ orbitals 
due to the pseudo-spin symmetry, which is realized commonly in the relativistic mean-field models. 

We found that the $B(GT;^{102}{\rm Sn}\to{}^{102}{\rm Sb})$ strength distribution is clearly different between 
non-relativistic and relativistic models 
while the low-lying $0^+$ and $1^+$ energy spectra  is similar to each other. 
By analyzing the compositions of wave functions and the expectation values of the residual interaction $V_{\rm res}$ in Eq. (\ref{eq:vpair}), 
we observe that the occupation of the $g_{7/2}$ orbital enhanced in the relativistic model 
breaks the $(L,S,T)=(0,1,0)$ coupling scheme of the two valence nucleons, and it brings about the substantial difference in the low-lying 
Gamow-Teller strength distributions. 

In conclusion, it is expect that the coupling in $(L,S,T)=(0,1,0)$ channel is strongly suppressed in the 
$^{100}{\rm Sn}+p+n$ nucleus due to the $g_{7/2}$ orbital, which is nearly degenerate to the $d_{5/2}$ orbital. 
The effect appears clearly in the low-lying $B(GT_-)$ distribution rather than in the energy difference  of $1^+_1$ to  $0^+_1$ states. 

Our three-body model developed here is a useful tool to investigate the structure of spherical core plus 
two nucleons systems not only with the non-relativistic but also with the relativistic models. 
As is found in the present study, it could be interesting to compare the results between 
non-relativistic and relativistic models with an attention to the characteristic features of the mean-field models. 
A systematic study of $T=0$ coupling effect for $N=Z$ odd-odd nuclei is one of the possible future works. 
Besides, an extension of our model for solution of the time-dependent Schr\"odinger (or Dirac) 
equation, as have been done in Refs. \cite{Oishi14,Maru}, 
provides a microscopic tool to study the decay dynamics of the valence nucleons. 
It will enable us to explore the effect of the neutron-proton correlation on the width of proton emission 
and also the possibility of deuteron emission from $^{102}$Sb nucleus.

\acknowledgements
We thank D. Lacroix, M. Grasso, and P. Schuck for helpful discussions. 
Y.T. acknowledges the grant received by the label P2IO. 
\appendix

\begin{widetext}

\section{Relativistic version of the model} \label{app:3brel}

\subsection{Single-particle states and two-particle basis}

The single-particle states in the relativistic EDF calculations are given by solutions of the single-particle Dirac equation. 
If the mean field is spherically symmetric, the single-particle wave function can be written as an eigenstate 
of the total angular momentum, 
\begin{eqnarray}
\varphi_{am}(\bvec r) = 
\left(\begin{array}{c}
\phi_{+,a}(r)\mathscr{Y}_{l_a^+j_am}(\theta,\phi) \\
i\phi_{-,a}(r)\mathscr{Y}_{l_a^-j_am}(\theta,\phi)  
\end{array}\right), 
\end{eqnarray}
where $l_a^+=l_a^-\pm 1$ for $j_a=l_a^+\mp 1/2$. Note that $+$ and $-$ distinguish 
the upper and lower components of a Dirac spinor. 
Thus it is labeled by the same quantum numbers as in non-relativistic case, $\{a,m\}=\{n_a,l_a^+,j_a,m\}$. 

The two-particle basis reads 
\begin{eqnarray}
\langle\bvec r_p,\bvec r_n|ab,JM\rangle &=& \sum_{m_am_b}\langle j_a m_a j_b m_b|JM\rangle
\langle\bvec r_p|\varphi^{(p)}_{n_al_a^+j_am_a}\rangle
\langle\bvec r_n|\varphi^{(n)}_{n_bl_b^+j_bm_b}\rangle
\nonumber \\
&=&
\sum_{m_am_b}\langle j_a m_a j_b m_b|JM\rangle
\left(\begin{array}{c}
\phi_{+,a}^{(p)}(r_p)\mathscr{Y}_{l_a^+j_am_a}(\theta_p,\phi_p) \\
i\phi_{-,a}^{(p)}(r_p)\mathscr{Y}_{l_a^-j_am_a} (\theta_p,\phi_p)
\end{array}\right)
\otimes
\left(\begin{array}{c}
\phi_{+,b}^{(n)}(r_n)\mathscr{Y}_{l_b^+j_bm_b}(\theta_n,\phi_n) \\
i\phi_{-,b}^{(n)}(r_n)\mathscr{Y}_{l_b^-j_bm_b}(\theta_n,\phi_n) 
\end{array}\right). 
\end{eqnarray}

The matrix element of the Hamiltonian is given by 
\begin{eqnarray}
&&H^{(J)}_{a'b',ab}
=
\langle a'b',J|H|ab,J\rangle \nonumber \\
&=&
\delta_{a'a}\delta_{b'b}(\epsilon^{(p)}_a + \epsilon^{(n)}_b) + 
\langle a'b',J|V_{\rm res}(S=0)|ab,J\rangle + \langle a'b',J|V_{\rm res}(S=1)|ab,J\rangle, 
\end{eqnarray}
where $V_{\rm res}(S=0)$ and $V_{\rm res}(S=1)$ corresponds to the first and the second term, respectively, 
in Eq. (\ref{eq:vpair}). 
This is diagonalized to get the eigenstate 
\begin{eqnarray}
|\Psi_{JM}\rangle = \sum_{ab}C^{(J)}_{ab}|ab,JM\rangle. 
\end{eqnarray}

\subsection{Pairing interaction and its matrix elements}
The residual interaction involves the projection onto spin-singlet and triplet. 
Since the spin operator $\bvec S$ satisfies the SU(2) algebra as the 
non-relativistic Pauli matrices, we give the relativistic version 
by simply replacing $\bvec\sigma$ by $\bvec\Sigma$, 
\begin{eqnarray}
\hat P_s=\frac{1}{4}-\frac{1}{4}\bvec\Sigma(p)\cdot\bvec\Sigma(n),\ 
\hat P_t=\frac{3}{4}+\frac{1}{4}\bvec\Sigma(p)\cdot\bvec\Sigma(n),  
\end{eqnarray}
which project a two-particle state on the singlet and triplet representations, 
respectively, of the SU(2) generated by $\bvec S=\frac{1}{2}\bvec\Sigma$. 

The matrix element of spin-singlet part of $V_{\rm res}$  reads 
\begin{eqnarray}
&&
\langle a'b',JM|V_{\rm res}(S=0)|ab,JM\rangle
\nonumber \\
&=&
\sum_{\eta,\eta'=\pm}
\int_0^{R_{\rm box}} dr\ r^2 g_s(r)
\phi^{(p)*}_{\eta,a'}(r)\phi^{(n)*}_{\eta',b'}(r)
\phi^{(p)}_{\eta,a}(r)\phi^{(n)}_{\eta',b}(r)
\nonumber \\
&&\times
\int d\Omega_p \int d\Omega_n\ 
\sum_{\lambda}Y_\lambda(\theta_p,\phi_p)\cdot Y_\lambda(\theta_n,\phi_n)
\nonumber \\
&&\times
\Bigl[\mathscr{Y}_{l_{a'}^\eta j_{a'}}(\theta_p,\phi_p)
\mathscr{Y}_{l_{b'}^{\eta'} j_{b'}}(\theta_n,\phi_n)\Bigr]_{JM}^\dagger
\left(\frac{1}{4}-\frac{1}{4}\bvec\sigma(p)\cdot\bvec\sigma(n) \right)
\Bigl[\mathscr{Y}_{l_a^\eta j_a}(\theta_p,\phi_p)
\mathscr{Y}_{l_b^{\eta'}j_b}(\theta_n,\phi_n)\Bigr]_{JM}, 
\end{eqnarray}
where 
\begin{eqnarray}
g_s(r)=V_0\left[1+x_s\left(\frac{\rho(r)}{\rho_0}\right)^{\alpha_s}\right]. 
\end{eqnarray}
From this we get 
\begin{eqnarray}
\langle a'b',JM|V_{\rm res}(S=0)|ab,JM\rangle
&=&
\frac{(-1)^{j_{a}-j_{a'}}}{8\pi}
\hat{j}_a\hat{j}_{a'}\hat{j}_b\hat{j}_{b'}
\left(\begin{array}{ccc}
j_a & j_b & J \\ 
-\frac{1}{2} & \frac{1}{2} & 0
\end{array}\right)
\left(\begin{array}{ccc}
j_{a'} & j_{b'} & J\\ 
-\frac{1}{2} & \frac{1}{2} & 0 
\end{array}\right)
\nonumber \\
&&\times
\sum_{\eta\eta'}
\delta_{l_a^\eta+l_b^{\eta'}+J,{\rm even}}\delta_{l_{a'}^\eta+l_{b'}^{\eta'}+J,{\rm even}}
(-1)^{l_a^\eta-l_{a'}^\eta}
\nonumber \\
&&\times
\int_0^{{R_{\rm box}}}r^2dr\ 
g_s(r)
\phi^{(p)*}_{\eta,a'}(r)\phi^{(n)*}_{\eta',b'}(r)
\phi^{(p)}_{\eta,a}(r)\phi^{(n)}_{\eta',b}(r)
\end{eqnarray}

The matrix element of the spin-triplet part of interaction is obtained in a similar way as 
\begin{eqnarray}
&&
\langle a'b',JM|V_{\rm res}(S=1)|ab,JM\rangle
\nonumber \\
&=&
\frac{(-1)^{j_a+j_b+j_{a'}+j_{b'}}}{4\pi}
\hat{j}_a\hat{j}_{a'}\hat{j}_b\hat{j}_{b'}
\sum_{\eta\eta'}
\sum_{L=|J-1|}^{J+1}\delta_{l_a^\eta+l_b^{\eta'}+L,{\rm even}}
\delta_{l_{a'}^\eta+l_{b'}^{\eta'}+L,{\rm even}}\hat{L}^2
\nonumber \\
&&\times
\left[\frac{(-1)^{j_b+1/2+l_b^{\eta'}}}{\sqrt{2}}
\left(\begin{array}{ccc}
j_a & j_b & J\\
-\frac{1}{2} & \frac{1}{2} & 0
\end{array}\right)
\left(\begin{array}{ccc}
L & 1 & J\\
0 & 0 & 0
\end{array}\right)
-
\left(\begin{array}{ccc}
j_a & j_b & J\\
-\frac{1}{2} & -\frac{1}{2} & 1
\end{array}\right)
\left(\begin{array}{ccc}
L & 1 & J\\
0 & -1 & 1
\end{array}\right)
\right]
\nonumber\\
&&\times
\left[\frac{(-1)^{j_{b'}+1/2+l_{b'}^{\eta'}}}{\sqrt{2}}
\left(\begin{array}{ccc}
j_{a'} & j_{b'} & J\\
-\frac{1}{2} & \frac{1}{2} & 0
\end{array}\right)
\left(\begin{array}{ccc}
L & 1 & J\\
0 & 0 & 0
\end{array}\right)
-
\left(\begin{array}{ccc}
j_{a'} & j_{b'} & J\\
-\frac{1}{2} & -\frac{1}{2} & 1
\end{array}\right)
\left(\begin{array}{ccc}
L & 1 & J\\
0 & -1 & 1
\end{array}\right)
\right]
\nonumber\\
&&\times
\int_0^{{R_{\rm box}}}r^2dr\ 
g_t(r)
\phi^{(p)*}_{\eta,a'}(r)\phi^{(n)*}_{\eta',b'}(r)
\phi^{(p)}_{\eta,a}(r)\phi^{(n)}_{\eta',b}(r)
\end{eqnarray}
where 
\begin{eqnarray}
g_t(r)=fV_0\left[1+x_t\left(\frac{\rho(r)}{\rho_0}\right)^{\alpha_t}\right]. 
\end{eqnarray}

\subsection{Decomposition into $S=0$ and $S=1$}

Since the matrix elements of the projector onto singlet is given by 
\begin{eqnarray}
&&\frac{1}{4}\langle a'b',JM|1-\bvec\Sigma(p)\cdot\bvec\Sigma(n)|ab,JM\rangle \nonumber \\
&=&
\sum_{\eta,\eta'}
\int_0^{R_{\rm box}} dr\ r^2\phi_{\eta,a'}^{(p)}(r)^*\phi_{\eta,a}^{(p)}(r)
\int_0^{R_{\rm box}} dr'\ r'^2\phi_{\eta',b'}^{(n)}(r')^*\phi_{\eta',b}^{(n)}(r')
\nonumber \\
&&\times
\int d\Omega_p \int d\Omega_n\ 
\bigl[\mathscr{Y}_{l_{a'}^\eta j_{a'}}(\theta_p,\phi_p)
\mathscr{Y}_{l_{b'}^{\eta'} j_{b'}}(\theta_n,\phi_n)\bigr]_{JM}^\dagger
\left(\frac{1}{4}-\frac{1}{4}\bvec\sigma(p)\cdot\bvec\sigma(n) \right)
\bigl[\mathscr{Y}_{l_a^\eta j_a}(\theta_p,\phi_p)
\mathscr{Y}_{l_b^{\eta'}j_b}(\theta_n,\phi_n)\bigr]_{JM}, 
\nonumber \\
\end{eqnarray}
the probability of $S=0$ is given as 
\begin{eqnarray}
P(S=0)
&=&
\frac{1}{2}\sum_{ab,a'b'}C_{a'b'}^{(J)*}C_{ab}^{(J)}(-1)^{j_b-j_{b'}}
\hat j_a\hat j_{a'}\hat j_b\hat j_{b'}
\nonumber \\
&&\times
\sum_{\eta,\eta'}
\delta_{l_{a'}^\eta l_a^\eta}\delta_{l_{b'}^{\eta'}l_b^{\eta'}}
\left\{\begin{array}{ccc}
j_a & j_b & J\\
l_b^{\eta'} & l_a^\eta & \frac{1}{2}
\end{array}\right\}
\left\{\begin{array}{ccc}
j_{a'} & j_{b'} & J\\
l_b^{\eta'} & l_a^\eta & \frac{1}{2}
\end{array}\right\}
\nonumber \\
&&\times
\int_0^{R_{\rm box}} dr\ r^2\phi_{\eta,a'}^{(p)*}(r)\phi_{\eta,a}^{(p)}(r)
\int_0^{R_{\rm box}} dr'\ r'^2\phi_{\eta',b'}^{(n)*}(r')\phi_{\eta',b}^{(n)}(r')
\end{eqnarray}

\subsection{Gamow-Teller transition}

It is assumed that the charge-exchange Gamow-Teller-type transition is induced by 
the coupling of the charge-changing axial-vector current of nucleon to the charged pion. 
The operator of this transition is given by 
\begin{eqnarray}
O(GT_\pm)=\sum_i\bvec\Sigma(i) t_\pm(i). 
\end{eqnarray}
Its matrix element from a neutron state $a$ to a proton state $a'$ is given by 
\begin{eqnarray}
\langle a'\|\bvec\Sigma\|a\rangle 
=(-)^{j_{a'}+3/2}\sqrt{6}\hat j_{a'}\hat j_a 
\sum_{\eta=\pm}\delta_{l_{a'}^\eta l_{a}^\eta}(-)^{l_{a}^\eta}
\left\{\begin{array}{ccc}
\frac{1}{2} & j_{a'} & l_a^\eta\\
j_a & \frac{1}{2} & 1
\end{array}\right\}
\int_0^{R_{\rm box}}dr\ r^2\phi_{\eta,a'}^{(p)*}(r)\phi_{\eta,a}^{(n)}(r)
\end{eqnarray}

\end{widetext}

\end{document}